\documentstyle[epsf,12pt]{article}
\newcommand{\al}{\alpha'}
\newcommand{\de}{\partial}
\newcommand{\be}{\begin{equation}}
\newcommand{\vp}{\varphi}
\newcommand{\ba}{\begin{eqnarray}}
\newcommand{\ea}{\end{eqnarray}}
\newcommand{\ee}{\end{equation}}

\newcommand{\we}{\wedge}

\newcommand{\lr}{\leftrightarrow}
\newcommand{\f}{\frac}
\newcommand{\s}{\sqrt}
\newcommand{\ti}{\tilde}
\newcommand{\ap}{\alpha}

\newcommand{\ddd}{\cdot\cdot\cdot}
\newcommand{\no}{\nonumber \\}
\newcommand{\la}{\langle}
\newcommand{\lb}{\rangle}
\newcommand{\ep}{\epsilon}

\newcommand{\ov}{\overline}

\begin{document}
\begin{titlepage}
\thispagestyle{empty}
\begin{flushright}
hep-th/0307083 \\
\end{flushright}
\bigskip
\bigskip
\begin{center}

\noindent{\Large \textbf{A Matrix Model Dual of \\
Type 0B String Theory
in Two Dimensions}}\\
\bigskip
\bigskip
\noindent{
          Tadashi Takayanagi\footnote{
                 E-mail: takayana@wigner.harvard.edu} 
 and \, Nicolaos Toumbas\footnote{
                 E-mail: nick@legendre.harvard.edu}}\\

\bigskip
{Jefferson Physical Laboratory\\
Harvard University\\
Cambridge, MA 02138}

\end{center}
\begin{abstract}
We propose that type 0B string theory in two dimensions admits a dual
description in terms of a one dimensional bosonic matrix model of a
hermitian matrix. The potential in the matrix model is symmetric with
respect to the parity-like $Z_2$ transformation of the matrix. 
The two sectors in the
theory, namely the NSNS and RR scalar sectors correspond to 
two classes of
operators in the matrix model, even and odd under the $Z_2$ symmetry
respectively. We provide evidence that the matrix model successfully
reconstructs the perturbative S-matrix of the string theory, and 
reproduces the
closed string emission amplitude from unstable D-branes.
Following
recent work in two dimensional bosonic string, 
we argue that the matrix model can be identified with the
theory describing N unstable D0-branes in type 0B theory. We 
also argue that type 0A theory is described in terms of the quantum
mechanics of brane-antibrane systems.
\end{abstract}
\end{titlepage}

\newpage

\section{Introduction}
\setcounter{equation}{0}
As is well known, two dimensional bosonic string theory admits a 
nonperturbative description in terms of a one dimensional matrix 
model, the
$c=1$ matrix model 
(for a comprehensive review of this duality see \cite{K,GM,Pr}). The
matrix model describes the quantum mechanics of an $N\times N$ 
hermitian matrix $\Phi$ in some potential.
On-shell closed string tachyon vertex operators
correspond to the `puncture' operator in the matrix model as follows
\ba
e^{iP(X_0 + \phi)}e^{2\phi}\  \lr \  \lim_{l\to 0}
\int dx e^{iPx}\mbox{Tr} e^{-l\Phi(x)}. \label{opbos}
\ea
Recently, a clearer understanding of this mysterious duality has emerged. 
As argued in references \cite{MV,KMS,MVT}, the $N\times N$ 
hermitian matrix 
$\Phi$
should be identified with the open string 
tachyon field $T$ on $N$ unstable D0-branes
in the bosonic string theory. Therefore, this duality is another example of
holographic open-closed string dualities. It fits nicely in the web of
holographic dualities such as Matrix theory \cite{BFSS} and the AdS/CFT
correspondence \cite{malda}.

In this paper we propose a similar description of the fermionic string in
two dimensions in terms of a one dimensional matrix model. In particular
type 0B string theory in two dimensions is described in terms of the
quantum mechanics of a hermitian $N\times N$ matrix $T$ with a potential
symmetric under $T \lr -T$: namely $U(T)=U(-T)$. The matrix model we have
in mind is the
model describing a collection of $N$ unstable 
D0-branes in this
theory\footnote{In flat ten dimensions, type 0 theory has doubled 
spectrum of D-branes \cite{BiSa,BG,KT}. However, as shown in \cite{FH,ARS}, 
there is 
only one type of spin structure allowed for boundary states in
$N=1$ super Liouville theory. Thus in the two dimensional 
fermionic string, we
only have one type of unstable D0-branes.}. 
The hermitian matrix $T$ is identified with the open string tachyon
field on the D0-branes. Thus we obtain another example of open-closed
string duality. 

The $Z_2$ symmetry of the potential in the fermionic case arises due to the
fact that the open string tachyon potential in the theory of D0-branes
obeys such symmetry. This symmetry is crucial in our analysis. The form of
the potential is such that a maximum at $T=0$ separates two stable minima
on each side. Unlike the bosonic case where the potential is believed to be
unbounded from below having only a metastable minimum, the vacua in the
fermionic case are stable. Both sides of the potential can be filled with
fermion eigenvalues
resulting in a fermi sea that is left/right symmetric. We propose that such
a symmetric ground state corresponds to the 0B vacuum. 
Thus unlike the case of the bosonic string, where only one side of the
potential is filled, the closed string vacuum in the fermionic case is
stable even non--perturbatively. 
In the bosonic case, the vacuum is non-perturbatively unstable
since tunneling effects can lead to particle loss on the unbounded side
(for recent discussions on the relation 
between non-perturbative effects \cite{Poi} in the matrix model and 
D-branes see \cite{Ma,AKK}).

Operators in the matrix model fall naturally into two sectors. Those
operators which are even under the $Z_2$ symmetry and those which are
odd. These operators describe disturbances of the fermi sea which are left
and right symmetric and antisymmetric respectively. 
We propose to identify each sector of operators of definite parity 
with the two
sectors of type 0B theory. The NSNS scalar admits a dual description in
terms of even operators in the matrix model while the RR scalar in terms of
odd operators. We provide evidence for such a correspondence by comparing
scattering amplitudes on the super Liouville side with matrix model
computations. We provide another check of the correspondence
by computing the closed string emission amplitude \cite{LLM,KMS,MVT,KLMS} 
during the process of open string tachyon 
condensation \cite{sbrane,S1,S2,OS,St,CLL,LNT,GS,MSY,GIR}. 
As we will see, this gives a precise relation between 
operators in the matrix model and closed string states in the NSNS 
and RR sectors.

In the more familiar example of the AdS/CFT correspondence, a
similar relation to eq. (\ref{opbos}) occurs between 
on-shell closed string states and gauge theory operators. 
Expectation values of closed string bulk fields appear as couplings of the
gauge theory operators. A similar picture seems to emerge in the matrix
model description of two dimensional string theory.
Indeed operators of the form (\ref{opbos}) in the theory of unstable
D0-branes couple naturally to the closed string tachyon. 

The dependence on
the tachyon field in the exponential in (\ref{opbos}) can be 
understood intuitively as follows. Such
operators in the matrix model cut loops of proper length $l$ in the dual
Riemann surface
\cite{MO,MS,K}. 
In the presence of an open string tachyon, a boundary interaction is added
to the worldsheet conformal field theory given by
$\int_{\de\Sigma}d\sigma T \sim lT$. The
exponential form of this interaction appears as 
in boundary string field theory (BSFT) \cite{Wi,GeSh,KuMaMo}. 

Motivated by such observations, we propose in the fermionic case (type 0B
theory) the following operator as `dual' to the
tachyon field in the NSNS sector 
\ba
\int dx e^{iPx}\mbox{Tr}\ e^{-lT^2(x)}. \label{opsup}
\ea
The Gaussian form is implied by world-sheet supersymmetry 
as is familiar in BSFT \cite{KuMaMo2,TTU}.
For the RR sector, we propose the odd function analogue 
\be
\int dx e^{iPx} \mbox{Tr}\ T(x) e^{-lT^2(x)} 
\ee
of (\ref{opsup}). These operators yield the correct leg factors that appear
in the scattering amplitudes of super Liouville theory.

The paper is organized as follows. In section 2, we review some useful
results in super Liouville theory. We describe how to obtain the type 0A
and type 0B theories in two dimensions and their spectra. 
We summarize the results for
closed string scattering amplitudes in these theories obtained in
\cite{DK}. 
We also study the decay of unstable
D0-branes in 0B theory. In section 3, we propose a matrix model description
for 0B theory in two dimensions. We explain how to obtain 
the string scattering
amplitudes from correlation functions in the matrix model with the correct
leg factors. We further illuminate the correspondence by identifying
decaying branes in string theory with rolling eigenvalues in the matrix
model and macroscopic loop operators in the matrix model with Euclidean 
D-branes. We also mention the matrix model dual of type 0A theory.
In section 4 we include a brief discussion. 

As we were finishing this work, we learned of related work 
to appear by
M. Douglas, I. Klebanov, D. Kutasov, J. Maldacena, E. Martinec 
and N. Seiberg
\cite{DKKMMS}. 
I. Klebanov has presented similar ideas
to ours in his talk at strings 2003 \cite{talk}.  

\section{$N=1$ Non-critical Closed String}
\setcounter{equation}{0}

The $N=1$ super Liouville theory  
is defined by the following action
\ba
S=\f{1}{2\pi}\int dz^2\int d\theta^2[D\Phi\bar{D}\Phi+2\mu_0 e^{b\Phi}],
\ea
where $\Phi$ is a superfield. 
The central charge is given by $c=\f{3}{2}(1+2Q^2)$, 
where $Q=b+\f{1}{b}$.
The conformal dimension of the primary operators 
$V_{\ap}=e^{\ap\phi}$ is given by
$\ap(Q-\ap)/2$. On the string theory side we always set
$\al=2$. 

The $N=1$ two dimensional string 
(or fermionic $\hat{c}=1$ non-critical string theory) 
\cite{P,DHK,DK} is obtained by setting $b=1, Q=2$ and adding a
world-sheet scalar field $Y$ together with its super partner $\psi$. 
The total central charge is canceled
by $b,c,\beta,\gamma$ ghosts ($c_{gh}=-15$). We will 
implicitly consider the
analytical continuation $Y=iX^{0}$ into two dimensional Minkowski space.
In this meaning, we also use the vector 
notation $X^{\mu}=(X^{0},\phi)$.

Since the Liouville term in the action is obtained by adding the tachyon
vertex operator, we consider type 0 theory (non-chiral GSO projection) in 
order not to project it out.
The field content in NSNS sector consists only of 
the `tachyon field' $T_{cl}$, which 
becomes massless due to Liouville dressing (for a review of 
quantum Lioville theory see \cite{SRV,Tes}).
Its vertex operator in the (-1) picture is 
\ba
V_{NSNS}=e^{-\vp}e^{iPX_0}e^{(\f{Q}{2}\pm iP)\phi},
\ea
where only one linear combination (of $\pm$ sign) is allowed for the wave 
function due to reflection at the Liouville wall; thus 
we can choose the $+$ sign \cite{Tes}.

In the RR-sector, we 
have two choices of GSO projection: type 0A and 0B. This fact is shown by
requiring the locality of OPEs as in the ten 
dimensional type 0A and 0B theories (for a review of 
ten dimensional type 0 string theory see \cite{Pol}). 
To be more explicit, let us
employ the bosonization of fermions $\psi_{\phi}+i\psi=e^{ih}$ ($h$ is the 
new boson). Then the physical R vertex in the 
$-\f{1}{2}$ picture is given by
\ba
V_{R(\ep)}=e^{-\f{1}{2}\vp}e^{\f{i}{2}\ep h}e^{iPX_0+(\f{Q}{2}-i\ep P)\phi}.
\ea
Note that for fixed $\ep$ the operator has 
only a left or a right-moving mode.
The total RR vertex can be constructed by left and right combination of 
the above
operators $V_{R(\ep_L)}\otimes V_{R(\ep_R)}$.
 The type 0A theory corresponds to the choice 
$(\ep_L,\ep_R)=(+,-),(-,+)$, while type 0B to $(\ep_L,\ep_R)=(+,+),(-,-)$.

Since the RR sector of type 0B theory is left-right symmetric, it has
a well-defined RR-scalar field $C$. Its field strength $F_{\mu}=\de_{\mu}C$ 
is divided into two (light-cone coordinate) components\footnote{ 
Due to reflection of the Liouville potential, 
only one linear combination is 
allowed; this is consistent with the fact that there is only 
one type of D-branes (electric), as we argue later.} $F_{+}$ and $F_{-}$. 
Indeed $F_{+}$ corresponds to $(\ep_L,\ep_R)=(+,+)$ and $F_{-}$ to $(-,-)$.
Note also that this RR-form is middle dimensional as RR-5 form in ten
dimensional type 0B
string and that there is no self-dual constraint in type 0 theory.
In this way type 0B theory includes two massless scalar fields $T_{cl}$ 
and $C$.

On the other hand, in type 0A theory the RR-vertex operator is like
$\sim e^{iP(X_{L}-X_{R})}$ and makes sense only at 
zero momentum (or constant
RR-flux). This is 
understood as follows. In two dimensional type 0A theory it is natural
to expect RR-1 form potential $C'$. Its 2-form field strength $F'=dC'$
should obey the equation of motion $d*F'=0$ and this allows only
constant flux (much like the massive IIA theory).

\subsection{Closed String Scattering Amplitudes and Leg Factors}

The three-point and four-point closed string scattering amplitudes were 
computed at tree level in
\cite{DK,AD}. 
In the NSNS sector the result is given by \footnote{We suppress an overall
delta function $\delta(\sum_i P_i)$ arising from energy conservation.} 
(where we speculate the results
for higher point functions very naturally)
\ba
A(P_1,P_2,\ddd,P_m)=\mu^{2-m}
\left[\prod_{j=1}^{m}\mu^{-iP_i}\cdot
\gamma(1+iP_j)
\right]S(\s{2}P_1,\s{2}P_2,\ddd, \s{2}P_m) \label{nsscat},
\ea
where $\gamma(x)=\Gamma(x)/\Gamma(1-x)$. The kinematic function $S$ 
of momenta
takes the same form as the one from the amplitude in bosonic string (so  
$S(P_1,P_2,\ddd,P_m)$ just represents the bosonic string counterpart).

For RR sector it is given by
\ba
A(P_1,P_2,\ddd,P_m)=\mu^{2-m}
\left[\prod_{j=1}^{m}\mu^{-iP_i}\cdot
 \gamma(\f{1}{2}+iP_j)\right]
 S(\s{2}P_1, \s{2}P_2,\ddd,\s{2}P_m)\label{rscat},
\ea
where we again find the same function $S$. Only scattering amplitudes
involving an even number of RR vertex operators are non-zero. Mixed
amplitudes have a similar structure: apart from the leg factors they are
similar to the bosonic string theory amplitudes as above. 

There is an essential difference between the scattering amplitudes
(\ref{nsscat}) and (\ref{rscat}). From the form of the leg factors in the
two equations, we see that the zero momentum NSNS tachyon decouples from
the spectrum while the zero momentum RR scalar does not \cite{DK}. This
fact was explained in \cite{DK}. The wavefunction of the zero momentum
tachyon is not normalizable, and therefore it decouples from the
spectrum. On the other hand, the wavefunction of the zero momentum RR
scalar is normalizable and need not decouple.

The results summarized above show
that the type 0 theory has essentially a similar 
perturbative structure as the 
bosonic string in two dimensions. They imply that the type 0 theory 
should also admit 
a dual matrix model description closely related to
$c=1$ matrix model, a point already made in \cite{DK}.

\subsection{Rolling Tachyon Computation}

In two dimensional type 0B string theory, we can consider 
the decay of unstable D0-branes.
The brane decay is described by an exact boundary conformal field theory 
(BCFT). We consider the case where\footnote{
It is also possible to consider tachyon fields of the form
$\ti{\lambda}\cosh(x^0/2)$ or $\ti{\lambda}\sinh(x^0/2)$ with the 
``Hartle-Hawking'' contour defining the analytic continuation \cite{LLM,KMS}.
We have to replace $\lambda$ with $\sin(\pi\ti{\lambda})$ or
$\sinh(\pi\ti{\lambda})$ respectively, (see \cite{S2}).}
 the rolling tachyon field is represented
by the $N=1$ boundary interaction (half-Sbrane) \cite{St,LNT,GS}
\be
\mu_B\int_{\de \Sigma}d\sigma\ \eta \psi e^{\f{1}{2}X_0},
\ee
where $\eta$ is the boundary fermion (as usual in BSFT formulation 
\cite{KuMaMo2}) of 
unstable D-branes. The D0-branes are described by the $(1,1)$ type 
degenerate
boundary state as argued in \cite{KMS,MVT,ZZ,FZZ}. In the $N=1$ 
super Liouville theory, the boundary states and one point functions
have been computed in \cite{FH,ARS} and below we use 
their results. A $(1,1)$ type D0-brane is  
localized in the 
strong coupling region, and the open string tachyon living on it has the
standard mass $m^2=-{1}/{2\al}$.

First let us consider the emission amplitude (see \cite{LLM,KMS} for
bosonic string case) 
of the NSNS tachyon field $T_{cl}$.
The closed string one point function of the vertex $e^{iEX_{0}}$ 
can be computed by analytical continuation as done in \cite{GS} 
\ba
A_{t}=\la e^{iEX_{0}} \lb =
-\f{1}{\s{2\pi}} 
(\s{2}\mu_B\pi)^{-2iE} \f{\pi}{\sinh(\pi E)}.
\ea
The Liouville part of the 
one point function of the vertex $e^{(iP+\f{Q}{2})\phi}$ on the (1,1) 
type D-brane
has been computed in \cite{FH}
\ba
A_{\phi}=\la e^{(iP+\f{Q}{2})\phi} \lb =-i(\mu_0\gamma(1)\pi)^{-iP} 
\f{\Gamma(iP) \sinh(\pi P)}{\Gamma(-iP)}.
\ea
Thus the emission rate of on-shell ($E=P$) closed string tachyon 
field is given by (up to a finite constant)
\ba
A_{NS}=A_{t}A_{\phi}
=\mu^{-iP} \f{\Gamma(iP)}{\Gamma(-iP)} e^{-iP\log\lambda}
, \label{rons}
\ea
where we renormalized the cosmological constant as 
$\mu\equiv\mu_0\gamma(1)\pi$ and also defined a 
parameter of the boundary interaction by $\lambda\equiv 2\pi^2\mu_B^2$. 

Next consider the emission amplitude of the RR-field $C$. 
In this case due to the spin field
of both $\psi_0$ and $\psi_1$ we get the factor $\cosh(\pi P)$ instead of
$\sinh(\pi P)$ in the timelike and spatial parts of the amplitude. In the 
final expression both cancel,  
and we end up with
\ba
A_{R}=\mu^{-iP} \f{\Gamma(\f{1}{2}+iP)}{\Gamma(\f{1}{2}-iP)}
 e^{-iP\log\lambda}.
\ea

One important common property of these emission amplitudes 
is the appearance of momentum dependent phase factors (leg factors) 
\ba
e^{i\delta_{NS}}&=&\mu^{-iP}\f{\Gamma(iP)}{\Gamma(-iP)}, \label{legns}\no
e^{i\delta_{R}}
&=&\mu^{-iP}\f{\Gamma(\f{1}{2}+iP)}{\Gamma(\f{1}{2}-iP)}.\label{legrr}
\ea
As in the bosonic string \cite{KMS,MVT}, this suggest that in a 
dual matrix model 
description, these phases may be interpreted
as standard leg factors relating matrix model operators and
closed string fields.

Note also the appearance of another phase factor $e^{-iP\log\lambda}$. In
half S-branes, the parameter $\lambda$ can be changed by a time
translation. Therefore, this phase can be interpreted as a time delay. This
phase can be also reproduced in the matrix model description \cite{KMS}.

\section{Matrix Model Description}
\setcounter{equation}0

We would like to propose that type 0B string theory in 
two dimensions is also
described by a one dimensional bosonic matrix model of a hermitian $N
\times N$ matrix $T$. Following the recent proposal of \cite{MV,KMS,MVT} the 
matrix $T$
is an open string tachyon field on $N$ unstable $D0$ branes. 
The ten dimensional type 0 theory contains double spectrum of 
D-branes: the so called electric branes $|-\lb$ and magnetic branes $|+\lb$ 
\cite{BiSa,BG,KT}.
However, in the two dimensional type 0 theory there is only one kind
of D-brane $|-\lb$ 
(we call it electric) as can be shown from the 
general analysis of boundary states in super Liouville theory \cite{FH}.
This implies that we have no fermionic fields on D-branes.
D0 branes in the theory are described by a
localized boundary state (corresponding to the $(1,1)$ degenerate state) 
for the Liouville direction $\phi$ times a Neumann
boundary state for the time direction $X^0$. The open string spectrum on
such branes consists of a tachyon field $T$ and a non-dynamical gauge field
$A_0$. There are no physical open string oscillator modes. 

A similar proposal for type 0A theory can be obtained
by considering a system of $N$ charged D0-branes and $N$ 
anti D0-branes. Even though 
the tachyon field becomes a complex scalar, the degrees of freedom 
after projection under the gauge group $U(N)\times U(N)$ are almost the 
same as in the case of unstable D0-branes in type 0B theory 
\footnote{We thank L. Motl
for discussions on this point.}.  

The action is naturally assumed to be $S=\beta\int dt L$ with the 
Lagrangian
\be
 L={1 \over 2}\mbox{Tr}(D_t T)^2 - \mbox{Tr}U(T),
\ee
and where
\be
D_tT=\partial_tT-i[A_0, T]
\ee
is the covariant derivative. $U(T)$ is the tachyon potential. The gauge
field $A_0$ acts as a Lagrange multiplier that projects onto $SU(N)$
singlet wavefunctions.

The Lagrangian above is of the same form as the Matrix model Lagrangian
used to describe the bosonic string theory with two important
differences. First the potential $U(T)$ is symmetric under the $Z_2$
transformation $T \rightarrow -T$, that is $U(T)=U(-T)$ 
(for example, see the review \cite{SR}). Second expanding
the potential near the quadratic maximum at $T=0$, we get
\be
U(T)=-{1\over 4\al}T^2 + O(T^4),
\ee 
that is the mass of the tachyon is given by $m^2=-1/2\al$ in string
units. This differs from the bosonic case by a factor of $1/2$. 
To make contact with the results in the bosonic matrix model (we follow
the convention in \cite{K})  
we set
$\al=1/2$. To compare with the results in super Liouville theory in which
$\al=2$, we rescale at the end energies and momenta by a factor of $2$: 
$(E,k)\rightarrow (2E, 2k)$.

Since the theory has an exact $Z_2$ symmetry, gauge invariant single trace
operators fall naturally into two categories. Operators which are even
under $T \rightarrow -T$ and operators which are odd. Correlation functions
in a state invariant under this $Z_2$ symmetry of an odd number of odd
operators are therefore zero. No such selection rule applies to even
operators.

The singlet sector of the Matrix model is exactly solvable due to the fact
that the $N$ eigenvalues of the matrix $T$ act as free non-relativistic 
fermions, and, therefore it can be described by Slater determinant
wavefunctions of $N$ variables. The eigenvalues of the matrix $T$ are
denoted by $\lambda_i$. The Hamiltonian is given by
\be
\left(\sum_{i=1}^Nh_i\right)\Psi(\lambda)=E\Psi(\lambda),
\ee
where
\be
h_i=-{1\over 2\beta^2}{d^2 \over d\lambda_i^2}+U(\lambda_i).
\ee
Since the single particle Hamiltonian is invariant under the parity
transformation $\lambda_i \rightarrow -\lambda_i$, the single-particle
eigenfunctions have definite parity. They are even or odd under
parity. The ground state wavefunction for a single particle bound state
will be even.  
The many particle ground state is obtained by filling\footnote{
Such a kind of model with two fermi seas was already considered before
in e.g. \cite{MO,DMW}.} the
first $N$ levels up to a fermi level $-\mu_F$ {\it on both sides of the
maximum}. The fermi level is measured from the local maximum of the
potential. We can choose $N$ even so that the ground state 
is invariant under
the $Z_2$ symmetry mentioned above. Then the fermi sea is symmetric.       

The double scaling limit is obtained by sending $\mu_F \rightarrow 0$ and
$\beta \rightarrow \infty$ keeping $\mu=\beta \mu_F$ fixed. The parameter
$\mu$ is proportional to $1/g_s$ and has to be kept large in perturbation
theory. This limit zooms near the top of the potential near which 
\ba
L=\mbox{Tr}[\f{1}{2}(\f{dT}{dt})^2+\f{1}{2}T^2+....],
\ea
We will often describe the theory in Euclidean 
time $x=-it$ and analytically continue.

The two sides of the fermi sea are independent in perturbation theory, that
is to all orders in the $g_s \sim 1/\mu$ expansion. The mixing occurs
through non-perturbative effects such as single eigenvalue
tunneling. Tunneling is suppressed by factors of order
$O(\exp(-\mu))$. Thus in the perturbative regime we can focus attention on
each side of the fermi sea separately. We call the two sides left(-) and
right(+). We essentially have two identical decoupled systems with a parity
operator that interchanges them. States in the system can be taken 
to be linear
combinations of 
\be
|\psi>=a|+>+b|->.
\ee
The parity operator is then the $2\times 2$ Pauli matrix $\sigma_1$ with
non-zero off-diagonal unit entries. Operators are naturally diagonal in
this basis. They act on each side separately without mixing them.
Even operators commute with parity. They are proportional to the $2\times
2$ identity matrix. Odd operators anticommute with parity. Therefore,
odd operators can be taken to be of the diagonal form 
\be
\int d\lambda f(\lambda) \rho(\lambda)\sigma_3, 
\ee
where we
have restricted $\lambda$ positive and $f(\lambda)$ is an odd function of
$\lambda$. Here $\rho(\lambda)$ is the density of eigenvalues
(or fermion bilinear). The ground state 
can be taken to be invariant under parity. In
computing correlation functions in the ground state amounts to taking a
trace over the $2\times 2$ diagonal matrices. Then we see that correlators
of an odd number of odd operators vanish. All other correlators are
identical to the standard case of the bosonic matrix model assuming we take
the ground state to be $(|\mu,+> + |\mu,->)/\sqrt{2}$, where the notation
indicates that both sides of the potential have been filled up to a fermi
level $-\mu$\footnote {The states
$|\mu,\pm>$ are not states of definite parity. They are good ground states
in perturbation theory. Non-perturbatively, only the left/right symmetric
state is a good ground state.}.

The second quantized Hamiltonian for a system of free fermions on each side 
is taken to be (see \cite{K,GK,SW,MO,MS})
\be
\hat{H}_{\pm}=\int_0^{\infty} dy \left[{1\over 2}{\partial  
{\Psi}^{\dag} \over
dy}{\partial \Psi \over dy} -{y^2 \over
2}\Psi^{\dag}\Psi+\mu(\Psi^{\dag}\Psi -{N \over 2})\right],
\ee 
where $y$ is the coordinate in the eigenvalue direction.
We follow the conventions of \cite{K}.
To proceed, it is convenient to bosonize the fermions. To this extent one
introduces new chiral fermionic variables $\Psi_L(t,\tau)$ and
$\Psi_R(t,\tau)$, where $\tau$ is defined through 
\be
v(y)={dy \over d\tau}=\sqrt{y^2 -2\mu},
\ee
the velocity of the classical trajectory of a particle at the fermi
level. Then the Hamiltonian is free and relativistic in terms of 
the new chiral
fermion fields up to interaction terms of order $1/v^2$ \cite{K}. These
interaction terms generate the $S$-matrix. In the large
$\tau$ region, $v$ is large, and therefore the relativistic 
Hamiltonian approximates well $\hat{H}$. 

The relativistic fermion $\Psi_L$ is bosonized \cite{K,GK,DJ,Polc,SW,KMS} as
\be
\Psi_L(\tau+t)|0>=e^{2\sqrt{\pi}X_L(\tau+t)}|0> \label{bX}.
\ee
A similar relation holds for the right movers as well. The field X lives on
the half line with Dirichlet boundary conditions at $\tau=0$. It can be
expanded as
\be
X(x,\tau)=\int dx e^{-iqx}\int dk \sin(k\tau)\tilde{X}(q,k).
\ee
The boundary conditions insure that the fermion 
current density vanishes at the
boundary at $\tau=0$. Finally, we note that the fermion density bosonizes
as follows \cite{K}
\be
\Psi^{\dag}_L\Psi_L + \Psi^{\dag}_R\Psi_R =-{1\over
\sqrt{\pi}}\partial_{\tau}X.
\ee

\subsection{Matching Operators and String States}

We begin our analysis by considering a typical macroscopic loop operator
in the standard bosonic theory: 
\be
\int dx e^{iPx}\mbox{Tr}\ e^{-lT(x)}. \label{macro}
\ee
 In the Lorentzian theory this
operator gives the following leg factor
\ba
\mu^{-iP}\f{\Gamma(2iP)}{\Gamma(-2iP)}, \label{legmat}
\ea
where we have translated the result for now in our $\al=2$ notation.
Naively this leads to crucially different results 
from (\ref{legns}). In fact, we get
poles at $iP=-1/2,-1,-3/2,...$ in eq. (\ref{legmat}), 
while in the NSNS sector
 leg factor poles occur at $iP=-1,-2,....$ and in the RR sector at
$iP=-1/2,-3/2,....$. These poles have an important 
physical meaning in the Euclidean
theory. They correspond to resonances in the Euclidean amplitudes 
due to the presence
of extra discrete states \cite{K}. These states are remnants of oscillator
modes of the string.

This observation suggests that the space of operators in $c=1$ 
bosonic matrix model can be divided naturally
into two sectors corresponding to the NSNS and RR sectors of type 0B theory. 
The operator (\ref{macro}) does not have definite parity under $T\lr
-T$. It is a mixture of both odd and even operators.
We propose that even operators in the matrix model
describe the NSNS sector and odd ones the RR sector.
This is a natural guess since
the model we propose has an exact $Z_2$ symmetry, the symmetry $T\lr -T$ of
the tachyon field. Indeed the closed string couplings on
unstable D-branes respect this symmetry as we will see below.

This choice is also natural from another perspective. By appropriately
amputating correlation functions of such operators, we
can obtain the 
scattering amplitudes in the Euclidean theory, as in the bosonic string
theory \cite{K}. As we mentioned
before correlators of an odd number of odd operators vanish in the
symmetric ground state. Scattering
amplitudes in the RR sector obey a selection rule: only scattering
amplitudes involving an even number of RR vertex operators are non-zero.

To study scattering amplitudes involving the 
NSNS sector tachyon we propose to study the small $l$ limit of correlators
of the following operator \footnote{In this section, we will suppress the
$2\times 2$ diagonal matrix accompanying the operators.}  
\ba
O_{NSNS}(q,l)=\int dx e^{iqx}\mbox{Tr}[e^{-lT(x)^2}].\label{nsop}
\ea
We will do all computations setting $\alpha'=1/2$ and rescale to the
$\alpha'=2$ notation in the end.
The standard `puncture' operator $P(q)$ should be the 
leading term in the small 
$l$
expansion of eq. (\ref{nsop}):
\be
P(q) \sim l^{-|q|/2}O_{NSNS}(q,l).
\ee 
The factor of $T^2$ in the exponential has an intuitive explanation. It is
based on an analogy with boundary string 
field theory (BSFT) \cite{Wi,GeSh,KuMaMo}. In BSFT 
the coupling of an 
unstable D-brane
to closed strings in the NSNS sector is proportional to 
the tachyon potential 
given by $e^{-\int d\sigma T^2}$ for superstrings \cite{KuMaMo2}
(and roughly $\sim e^{-\int d\sigma T}$ for the bosonic string). 
The same argument can be generalized for the type 0A case 
(constructed from $D0-\bar{D}0$ system).
We have to replace $e^{-lT(x)^2}$ with $e^{-lT\ov{T}(x)}$ 
in (\ref{nsop}) (--for the BSFT description of brane-antibrane systems 
see \cite{TTU}).

Upon bosonization (\ref{nsop}) becomes
\ba
O_{NSNS}(q,l) \sim \int dx e^{iqx}\int d\tau
e^{-ly^2(\tau)}\partial_{\tau}X\sim i\int_{-\infty}^{\infty}dk F(k, l)k
\tilde{X}(q, k) \label{bosonized}
\ea
where
\begin{equation}
F(k, l)=\int d\tau e^{-ly^2(\tau)} 
\cos(k\tau).
\ee
Evaluating $F(k,l)$ with $y(\tau)=\sqrt{2\mu}\cosh(\tau)$, we obtain
\be
F(k,l)={e^{-l\mu}\over 2}K_{ik/2}(l\mu)
\ee
where $K$ is a modified Bessel function.
In the small $l$ limit this function becomes
\be
F(k,l)\rightarrow {\pi \over 4 \sin({ik\pi \over
2})}\left((l\mu/2)^{-ik/2}{1 \over \Gamma(-ik/2+1)}-cc\right) 
\label{small}.
\ee

In computing $m$-point Euclidean correlators, each such 
operator will be connected to
the rest of the Feynman graph by the scalar field propagator 
$1/(q^2 +k^2)$. We can now perform the $k$-integral in (\ref{bosonized}) 
as in \cite{K}. The contour of intergration is deformed in a special way. 
For the first
term in (\ref{small}), we deform the integration so as to pick up the 
residue of the propagator pole at
$k=i|q|$ while for the second term at the pole $k=-i|q|$ \cite{K}. This
procedure gives the amputated on-shell Euclidean amplitude for $m$ X quanta
times a factor for each external leg (up to numerical factors)
\be
(l\mu/2)^{|q|/2}\Gamma(-|q|/2).  
\ee
This `leg factor', reproduces the correct pole structure of scattering
amplitudes in the Euclidean string theory due to NSNS discrete states.
In choosing the operator $O(q,l)$, we allow freedom to multiply it by a
smooth function of $|q|$. Thus we can multiply by a factor of
$1/\Gamma(|q|/2)$ (since this is a smooth function). This amounts to
smearing the local operator in position space. 

Translating into the $\alpha'=2$ notation amounts to rescaling the energies
$q$ by a factor of $2$. The Lorentzian theory is obtained by analytic
continuation $|q| \rightarrow -iP$.
We thus obtain the leg factor
\be
\mu^{-iP}{\Gamma(iP) \over \Gamma(-iP)}.
\ee
This agrees with (\ref{legns}).

Thus in the small $l$ limit, correlators of $m$ properly normalized 
NSNS operators take the form
\be
\mu^{2-m}\prod_{j=1}^m \left(\mu^{-iP_j}{\Gamma(iP_j) \over
2\Gamma(-iP_j)}\right)S'(\s{2}P_1,...,\s{2}P_m) \label{scats}
\ee  
with $\mu^{2-m} S'$ 
being the usual bosonic matrix model scattering amplitude. Since
there is a well-known
agreement between the bosonic string theory and the matrix model, the
equivalence with the superstring follows. 
In particular, $S'$ takes the form \footnote{In our conventions with
$\alpha'=2$, the leg factors in the bosonic theory are given by 
$\Gamma(i\sqrt{2}P)/\Gamma(-i\sqrt{2}P)$.} \cite{K}
\be
S'=\prod_{j=1}^m(i\sqrt{2}P_j)S,
\ee
where $S$ is the bosonic kinematic part appearing in
eq. (\ref{nsscat}). After rescaling the momenta as implied by
eq. (\ref{scats}), the result agrees exactly with the scattering amplitude
(\ref{nsscat}).  
Notice again that our result implies a
rescaling of the bosonic string momenta ($\alpha'=2$) by a factor of
$\sqrt{2}$ to get to the superstring. This difference arises due to the
difference in mass of the open string tachyon field by the same factor. 
In a previous section we have seen
that apart from leg factors, the kinematic parts of the super Liouville
amplitudes are reproduced from the bosonic ones up to the same
rescaling of the momenta \cite{DK}.

For the RR sector we need to choose an odd operator.
To reproduce the 
correct leg factor (\ref{legrr}), the simplest choice seems to be 
the operator
 \be
O_{RR}(q,l)= \int dx e^{iqx}\mbox{Tr}\ T e^{-lT^2}. \label{RRop}
\ee
This operator (\ref{RRop}) is consistent with the known
RR coupling\footnote{In the same 
way we can discuss  
operators for the RR sector in the two dimensional type 0A theory. 
However, we 
are left with no candidates since now gauge invariant operators 
must be even under the $Z_{2}$ symmetry.
This is consistent with the fact that there is no dynamical RR field
in the type 0A theory.}; the RR field $C$
couples to the open string tachyon as $\int C\we dT~ e^{-T^2}$ 
in BSFT \cite{KuMaMo2,TTU}. Roughly,
the operator (\ref{RRop}) should be dual to the field strength of $C$ 
(see also the discussions below).

After bosonization, we get the
following wavefunction $F(k,l)$
\begin{equation}
F(k, l)=\int d\tau y e^{-ly^2} \cos(k\tau).
\end{equation}
Since $y=\sqrt{2\mu}\cosh\tau$ 
\ba
F(k, l)&=&\sqrt{2\mu}e^{-l\mu}\int d\tau \cosh(\tau) e^{-l\mu \cosh(2\tau)} 
\cos(k\tau)\\
&=&\sqrt{\mu \over 2}e^{-l\mu}\int d\tau e^{-l\mu \cosh(2\tau)} 
\left(\cos[(-i+k)\tau]+\cos[(i+k)\tau]\right).
\ea
Evaluating the integral we obtain
\begin{equation}
F(k, l)={1 \over 2}\sqrt{\mu \over 2}e^{-l\mu} 
\left(K_{(1+ik)/ 2}(l\mu)+K_{(-1+ik)/2}(l\mu)\right).
\end{equation}
Now consider the small $l$ limit. In this limit, we have
\begin{equation}
 K_{(1+ik) / 2}(l\mu) \rightarrow {\pi \over 
 2\sin ((ik+1)\pi/2)}I_{(-ik-1)/2}(l\mu)
\end{equation}
and
\begin{equation}
K_{(-1+ik) / 2}(l\mu)\rightarrow {-\pi \over 2\sin ((ik-1)\pi/2)}
I_{(ik-1)/2}(l\mu). 
\end{equation}
The other terms can be dropped as they are smaller by a factor of $l$.
Finally we obtain
\begin{equation}
F(k,l)=\sqrt{\mu \over 2}(l \mu/2)^{-1/2} 
\left[{\pi \over 4\sin((ik+1)\pi/
2)}(l\mu/2)^{-ik/2}{1 \over \Gamma[(-ik-1)/2 +1]} +cc\right].
\end{equation}

Now let us perform the momentum $k$ integral. We deform the integration as
in \cite{K}. For the first term, we deform
the integration to pick the pole $k=i|q|$ from the $X$-propagator while for
the complex conjugate the opposite pole. So after the $k$ integral we get
the following leg factor (up to numerical factors)
\begin{equation}
\sqrt{\mu \over 2}(l\mu/2)^{(|q|-1)/2}\Gamma[(1-|q|)/2].
\end{equation} 
The leg factor reproduces the correct pole structure arising from the RR
discrete states.
As before we are free to multiply by a smooth function of $q$. Thus we can
divide by a factor of $\Gamma[(1+|q|)/2]$.
The puncture operator is obtained in the small $l$ limit as follows
\be
P(q)\sim l^{(-|q|+1)/2}O_{RR}(q,l). 
\ee
We then have to rescale $q$ by a factor of 2, and finally continue to the 
Lorentzian
theory. We end up with (in the $\al=2$ notation) the following leg factor
\begin{equation}
\mu^{-iP}{\Gamma(1/2+iP) \over \Gamma(1/2 -iP)}
\end{equation}
in agreement with (\ref{legrr}).

To obtain the correct scattering amplitudes of RR scalars
eq. (\ref{rscat}), we need to relate the RR vertex operator $V_{RR}$ to the
odd operator (\ref{RRop}) by an additional factor of momentum:
$|q|V_{RR}\sim O_{RR}$. This factor arises because the operator
(\ref{RRop}) is directly related to the gauge invariant 
field strength $dC$
rather than $C$ itself, and reflects the non-decoupling of the zero momentum
scalar discussed in section 2. The scattering amplitude (\ref{rscat}) is
reproduced if the leg factors in eq. (\ref{scats}) are replaced by the
momentum dependent factors
\begin{equation}
{(\mu^{-iP_j}/ 2iP_j)}{\Gamma(1/2+iP_j) \over \Gamma(1/2 -iP_j)}.
\end{equation}

The considerations of this section imply that the perturbative
structure of the S-matrix is universal in two dimensional string theory,
bosonic and fermionic. 
Even though quite significant in the Euclidean theory (since they contain
physical information), the leg factor
phases appearing in the Lorentzian S-matrix amplitudes do not affect
scattering cross-sections. In fact, they can be removed by a unitary
transformation on the states of the theory. Non-perturbatively, we should
expect a different picture to emerge. In the bosonic case, non-perturbative
effects such as eigenvalue tunneling can spoil unitarity of the
S-matrix and the stability of the vacuum. One typically expects the open
string tachyon potential to be unbounded from below. In the fermionic case
however, worldsheet supersymmetry implies that the potential is bounded
from below, and the closed string vacuum is stable. We do not expect
non-perturbative violations of unitarity of the S-matrix in this case.

\subsection{Rolling Eigenvalues and Brane Decay} 
Our considerations above suggest that we can 
identify decaying D0-branes in
type 0B theory with rolling eigenvalues in the matrix model as in
\cite{KMS,MV}. 
A decaying brane
results in a coherent state of closed strings \cite{LLM}. Similarly a
rolling eigenvalue down the potential results in a coherent state for the
bosonic field $X$ of the form (\ref{bX}) \cite{KMS}. 
The relation of this field to the string theory
scalar fields is
non-local. Our considerations in the previous section suggest that in
momentum space the relation is just a momentum dependent phase.  

In the case we consider, we can obtain coherent states on both sides of the
fermi sea by considering rolling of eigenvalue pairs. 
Since the NSNS sector is described by operators even under
parity $T\lr -T$, one particle states of the NSNS tachyon correspond to
symmetric one-particle states of $X$ of the form
\be
|k>_{NSNS}=e^{i\delta(k)_{NS}}(|k,+>+|-k,->)/\sqrt2.
\ee 
Such a disturbance of the fermi sea is left/right symmetric. 
Similarly for the RR scalar we get
\be
|k>_{RR}=e^{i\delta(k)_{R}}(|k,+>-|-k,->)/\sqrt2. \label{RRstate}
\ee 
Since the RR sector is described by odd operators, the expectation value of
these in the symmetric ground state and an odd state of the form
(\ref{RRstate}) is non-zero. The phases are the leg factors
eq. (\ref{legns}). The time delay in the matrix model is reproduced by
considering classical trajectories of eigenvalues that start away from the
local maximum \cite{KMS}.

\subsection{Macroscopic Operators and Open String Partition Function}  
Before finishing this paper, let us consider an interesting relation between
the annulus amplitude of Euclidean unstable 
D0-branes and the two point function of macroscopic loop operators of the
form (\ref{nsop}). Such D-objects are defined by Neumann boundary
conditions along the  
Liouville direction, and Dirichlet conditions for the Euclidean time
direction $X^0$. From the point of view of the Lorentzian theory, these
objects are D-instantons. In a recent paper \cite{Ma}, 
it is proposed that such 
branes in the bosonic string are described by macroscopic loop operators of
the form (\ref{opbos}).
Here we would like to check this correspondence for our two dimensional 
fermionic string case. The amplitude describes NSNS closed string exchange.

After integration of the closed string channel modulus, the annulus
partition function is given by 
\ba
Z=\int dP \int dE \f{e^{iP(x-x')}\cos(2\pi P\sigma)\cos(2\pi P\sigma')}
{\sinh^2(\pi E) (E^2+P^2)},\label{app}
\ea
where the parameter $\sigma$ is related to the boundary cosmological 
constant by $\mu_{B}^2=\f{2\mu_0\sinh^2(\pi \sigma)}
{\cos(\pi b^2/2)}$ \cite{FH}. 

Define the Laplace transform of the 
operator\footnote{Here we consider $|+\lb$
branes.
For $|-\lb$ branes, we need to put an extra factor $e^{l\mu}$
in the definition of dual operator.} (\ref{nsop}) by
\ba
W(\sigma,q)=\int \f{dl}{l}e^{-\mu_B^2 l}O_{NSNS}(k,l).
\ea
Though we need a renormalization of both $\mu_B$ and
$\mu$, we can cancel it by scaling $l$.
Then by using the formula
\ba
\int^\infty_{0}\f{dl}{l}e^{-\mu l \cosh(2\pi\sigma)}K_{ik}(\mu l)
=\f{\pi\cos(2\pi\sigma k)}{k\sinh(\pi k)},
\ea
we can see that the 
two macroscopic loop correlator 
$\int dq e^{iq(x-x')}\la W(\sigma,q)W(\sigma,-q)\lb$ 
is equal to 
the annulus amplitude (\ref{app}).
Note that the boundary cosmological constant
leads to the term $\mu_{B}^2\int_{\de\Sigma} d\sigma e^{\phi}$ 
in the superstring case. The two-point function can be obtained using eq. 
(3.18) and (3.20). It would be interesting to obtain such a correspondence
for the RR operators as well.       

\section{Conclusions}
In this paper we explored holographic dual matrix models of 
two dimensional type 0 theory. We mainly 
focused our analysis on the seemingly
simpler case of type 0B theory. The 
tachyon field on (infinitely) many unstable D-branes plays the role of the
elementary matrix field in the matrix model. In the fermionic string the
tachyon potential becomes left-right symmetric and the theory admits a
$Z_2$ symmetry. 
We checked our proposal for the duality by comparing tree level scattering 
amplitudes of closed strings with their matrix theory counterparts.
The dual matrix model
is the same as that of the bosonic string in the double scaling limit.
However, the exact $Z_2$ symmetry of the fermionic model hints at the
existence of 
a stable vacuum even at the non-perturbative level.
It would be an exciting future direction to compute non-perturbative
quantities in the symmetric matrix model and examine their 
implications for the
fermionic string.
Finally, it would be interesting to obtain a precise description of type 0A
in terms of a matrix model of a complex field.

\bigskip

\begin{center}
\noindent{\large \bf Acknowledgments}
\end{center}

We would like to acknowledge useful conversations with D. Bak, H.C. Cheng,
K. Hosomichi,  
I. Klebanov, I. Low,
S. Minwalla, L. Motl, A. Strominger, S. Terashima and S. Wadia. 
This work is supported
in part by DOE grant DE-FG02-91ER40654.

\end{document}